\documentclass[]{interact}

\usepackage[caption=false]{subfig}

\usepackage{graphicx}
\usepackage{hyperref}
\usepackage{xcolor}
\usepackage{xspace}
\usepackage{braket}
\usepackage{amsfonts,amsmath,amssymb}
\usepackage[T1]{fontenc}
\usepackage{textcomp}
\usepackage[utf8]{inputenc}
\usepackage{xr}
\usepackage{commath, etoolbox, ifthen, letltxmacro, microtype, xparse}

\usepackage{dcolumn}

\usepackage[numbers,sort&compress,merge]{natbib}
\bibpunct[, ]{[}{]}{,}{n}{,}{,}

\theoremstyle{plain}

\theoremstyle{definition}

\theoremstyle{remark}

\newcommand{\tripS}{\ensuremath{3s\,^3S^o_1}\xspace}%
\newcommand{\ordsim}{\mathord{\sim}}
\newcommand*{\ie}{i.\,e.}%
\providecommand*{\etal}{et al.\xspace}%
\newcommand*{\degree}[1]{\ensuremath{\ifx\\#1\\\else\xspace#1\,\fi^\circ}\xspace}%

\AtBeginDocument{%
	\LetLtxMacro\autoreforig\autoref%
	\RenewDocumentCommand{\autoref}{som}{%
		\IfBooleanF{#1}{%
			\hyperref[#3]%
		}{%
			\autoreforig*{#3}\IfValueT{#2}{\nobreak}\xspace\IfValueT{#2}{#2}%
		}%
	}
}

\begin{document}

\articletype{}

\title{Laser ionization detection of O($^3P_j$) atoms in the VUV; application to photodissociation of O$_2$}

\author{
\name{Xu-Dong Wang\textsuperscript{a}, David H. Parker\textsuperscript{a}, Sebastiaan Y.T. van de Meerakker\textsuperscript{a}, 
Gerrit C. Groenenboom\textsuperscript{a}, Jolijn Onvlee\textsuperscript{a}\thanks{CONTACT Jolijn Onvlee. Email: 
J.Onvlee@science.ru.nl}}
\affil{\textsuperscript{a}Institute for Molecules and Materials, Radboud University, 
	Heijendaalseweg 135, 6525 AJ Nijmegen, The Netherlands}}

\maketitle

\begin{abstract} 
Detection of nascent O($^3P_j$, $j=2,1,0$) atoms using one-photon resonant excitation to the \tripS 
state at $\ordsim 130$~nm followed by near-threshold ionization, \ie, 1 + 1' resonance enhanced multi-photon ionization 
(REMPI), has been investigated. The aim was to achieve low ion recoil, improved sensitivity, and reliable angular momentum 
polarization information, with an as simple as possible laser setup.  An efficient 1 + 1' scheme has been found where the VUV 
light for the first step 1 is generated by difference frequency ($2\omega_1 - \omega_2$) VUV generation by four wave mixing in 
Kr gas, and the ionization step 1' uses $2\omega_2$ around 289 nm. The presented scheme induces 9 m/s 
recoil of the O$^+$ 
ion using a two-dye laser system, and zero recoil should be possible by generating 302 nm radiation 
with a third dye laser.  While this approach is much more sensitive than a previous 1 + 1' scheme using 
212.6 nm for the 1' step, we found that the relatively intense radiation of the $2\omega_2$ beam does 
not saturate the 1' step. 
In order to test the ability of this scheme to accurately determine branching ratios, fine structure yields, and angular 
distributions including polarization information, it has been applied to O$_2$ photodissociation around 130 nm with subsequent 
O($^3P_j$) fragment detection.
\end{abstract}


\section{Introduction}

Ground-state oxygen atoms, O($^3P$), which are highly reactive due to two unpaired electrons, play an 
important 
role in various chemical environments. 
In the atmosphere, photodissociation of O$_2$ leads to the formation of O($^3P$) atoms, 
which can rapidly react with O$_2$ to form ozone, or with ozone to form two O$_2$ 
molecules~\cite{Chapman:PM10:369}. In the interstellar medium, collisions between O($^3P$) and helium 
or atomic and molecular hydrogen have been identified as an important cooling 
process~\cite{Lique:MNRAS474:2313}. In oxygen-containing plasmas, O($^3P$) atoms are formed that play 
an important role in the chemistry in these complex systems~\cite{Jeong:JPCA104:8027}.  
To better understand these processes, high-resolution lab-based experiments have to be performed. 

For these experiments, it is crucial to be able to detect O($^3P$) atoms in an efficient and 
sensitive way. Traditionally, laser induced fluorescence (LIF) was often used to detect O($^3P$) by 
exciting the atoms to the $3p\ ^3P$ state using two 226 nm photons and detecting the $3p 
\rightarrow 3s$ fluorescence at 845 nm \cite{Bischel:CPL82:85}.
Later, 2 + 1 resonance enhanced multiphoton ionization (REMPI) often became the preferred method of 
choice, especially in combination with detection techniques such as velocity-map imaging (VMI) 
\cite{Eppink:RSI68:3477}. This 2 + 1 REMPI scheme (\autoref[a]{fig:schemes}) uses 226 nm light to 
first excite O($^3P$) to 
the $3p \ ^3P$ state by absorption of two photons, followed by the absorption of an additional photon 
to ionize the atoms \cite{Bamford:PRA34:185}. This REMPI scheme is easy to use, since only one 
dye laser is needed, with a wavelength that is commonly produced. It is moreover a very sensitive 
detection technique.

However, in 2 + 1 REMPI, the three-photon energy is $>2$~eV higher than the ionization energy 
of the O($^3P$) atoms, resulting in excess energy. Nearly, but not all, of this excess energy appears 
in the electron kinetic energy. Momentum conservation results in a slow recoil of the O$^+$ partner, 
which causes recoil of $1 \cdot 10^{-4}$~eV, or 34~m/s \cite{Suits:RSI89:111101}. While small, this 
recoil velocity spoils the resolution of advanced molecular dynamics experiments, particularly those 
combining controlled molecular beams \cite{Meerakker:CR112:4828} and final state-selective detection 
by VMI \cite{Eppink:RSI68:3477}.

In this article we focus on improving a previously used \cite{Lambert:JCP121:10437} laser ionization 
method for detection of O atoms using vacuum ultraviolet (VUV) resonant selection of O($^3P_j$) with 
$j=2,1,0$, followed by near-threshold ionization to decrease the O$^+$ ion recoil. This 1 + 1' 
photoionization scheme (\autoref[a]{fig:schemes}) makes use of 130 nm VUV radiation to excite the 
O($^3P$) atom to the \tripS 
state, and $\ordsim 289$~nm UV light to subsequently ionize the atoms slightly above the threshold at 
302 nm. 
The greatly reduced recoil velocity of 
the O$^+$ ion paves the way towards high-resolution detection of O($^3P$). Besides recoil, other 
critical aspects of laser ionization methods include sensitivity, purity of state selection, product 
$m$ state polarization dependence, and ease or expense of use. All these aspects of the 1 + 1' 
detection scheme will be discussed here.

In the past, several studies have used 130 nm radiation for exciting O($^3P$) atoms to the \tripS 
state prior to detection. It is, for instance, possible to detect laser-induced fluorescence after 
this 
excitation step \cite{Dobele:APB39:91}. Another method uses a second laser with a wavelength of 305 nm 
to excite the intermediate $3s$ state to high-lying Rydberg states which are then field ionized in a 
time-of-flight O-atom Rydberg tagging experiment \cite{Lin:JCP119:251, Jones:RSI79:123106}. This 
results in a high temporal and thereby kinetic energy resolution, and a negligible recoil velocity. In 
general, the lifetime of the Rydberg state used limits the signal levels and the combination of 
imaging with
Rydberg tagging is not ideal due to conflicting experimental 
needs~\cite{Cruse:JCP121:4089}. However, the Rydberg tagging technique has been combined successfully 
with sliced VMI~\cite{Cruse:JCP121:4089} and this presents a suitable approach for high-resolution 
detection of O($^3P$).

Lambert \etal, on the other hand, employed a 1 + 1' REMPI method for photoionizing O($^3P$) atoms via 
the intermediate \tripS state \cite{Lambert:JCP121:10437}. In this pioneering study, two-photon 
resonance-enhanced difference 
frequency four-wave mixing ($2\omega_1 - \omega_2$) (\autoref[b]{fig:schemes}) was used to 
generate tunable VUV radiation 
in the 120-132 nm 
region. Their work focused on photodissociation of O$_2$ (\autoref[c]{fig:schemes}) via the $E(v=0,1)$ 
states at 124 and 120 nm, 
respectively, which have very large cross sections compared to excitation around 130 nm.  By 
using two different $\omega_2$ laser beams they generated two VUV wavelengths for 
dissociation via the $E$ state, and O($^3P$) excitation followed by ionization at 212.56 nm (i.e., the 
radiation used for excitation at 2$\omega_1$). In the 130 nm region, the low signal to background 
ratio did not allow full analysis of their O($^3P$) data. 

Our work follows this study of Lambert \etal \cite{Lambert:JCP121:10437}. We show that the use of a 
more intense laser beam with a longer wavelength for the ionization step increases the sensitivity 
sufficiently, while at the same time creating less ion recoil. We also apply our low 
recoil 1 + 1' method to VUV photodissociation of O$_2$ with O($^3P$) fragment detection, and test the 
ability of our 1 + 1' scheme to accurately determine branching ratios, fine structure yields, and 
angular distributions including polarization information. 

\begin{figure}
	\centering
	\includegraphics[width=0.5\columnwidth]{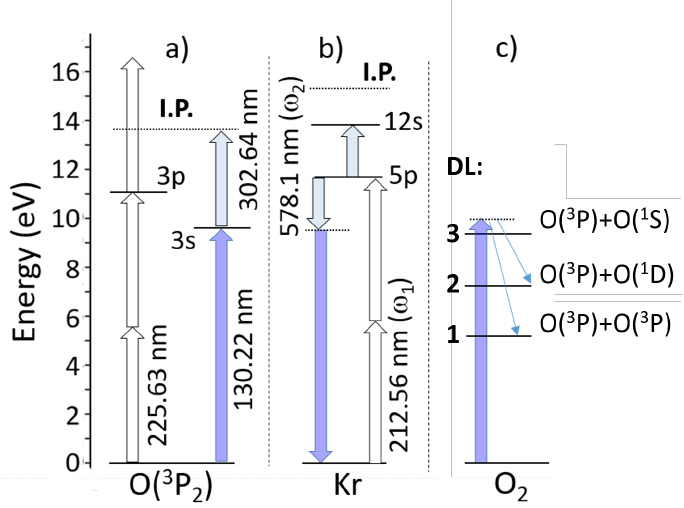}%
	\caption{Overview of laser methods used in this study. a) Standard 2 + 1 (left) and threshold 1 
		+ 1' (right) REMPI schemes for O($^3P_2$) detection. b) Difference frequency four wave mixing 
		scheme for 
		generating 130.22 nm light used in the 1 + 1' scheme. Small changes in the $\omega_2$ 
		wavelength 
		(see \autoref{fig:spectrum}) are used for detection of O($^3P_{j=0,1}$). c) Energetics of 
		O$_2$ 
		photodissociation around 130 nm. DL 1, 2, and 3 
		refer to the (O($^3P$) + O($^3P$)), (O($^3P$) + O($^1D$)), and (O($^3P$) + O($^1S$)) 
		dissociation 
		limits, respectively.}
	\label{fig:schemes}
\end{figure}

\section{Experimental methods}
In this study, we used a standard molecular beam apparatus with time-sliced VMI detection.
A supersonic molecular beam of neat O$_2$ was generated by a Nijmegen pulsed valve 
\cite{Yan:RSI84:023102}, 
using a backing pressure of 2.5 bar and a repetition rate of 10 Hz. A pulsed discharge in O$_2$ 
created O($^3P$) atoms \cite{Farooq:PCCP16:3305} that we used for testing the overall detection 
sensitivity. For the O$_2$ 
photodissociation study, however, the discharge was switched off. The skimmed molecular beam was 
intersected by photolysis and probe lasers that were perpendicular to the molecular beam propagation 
direction. The O($^3P$) atoms resulting from the discharge or photodissociation of O$_2$ were 
state-selectively ionized and the resulting ions were extracted perpendicular to the molecular beam 
propagation direction and focused onto a position-sensitive 
detector. This detector 
consisted of a microchannel plate (MCP) assembly coupled to a phosphor screen. A fast high-voltage 
switch with a pulse width of 100 ns was employed to gate the central slice of the atomic oxygen 
product's Newton sphere. Analysis of the images using the FINA program 
\cite{Thompson:JCP147:013913} 
indicated a 
typical $30\%$ degree of slicing. The VMI image was recorded by a CMOS camera and event counting was 
performed 
during the data acquisition. The final images were accumulated over $\ordsim 100\ 000$ laser shots. 
Conversion from pixel to m/s was obtained from an O($^3P_2$) image from O$_2$ photodissociation at 
225.63~nm.

The 130 nm radiation used for exciting O($^3P$) atoms to the \tripS state was generated by difference 
frequency mixing 
(2$\omega_1$- $\omega_2$) in krypton (Kr) gas as the nonlinear medium, where the $\omega_1$ and 
$\omega_2$ frequencies were generated by two dye lasers, which were simultaneously pumped by a 532 nm 
laser beam
from a single Nd:YAG laser operated at 10 Hz. The $\omega_1$ and $\omega_2$ laser beams were spatially 
and temporally overlapped and focused into a stainless-steel gas cell that was filled with 20 mbar of 
Kr. The VUV 
radiation, together with the $\omega_1$ and $\omega_2$ generation beams, passed through a MgF$_2$ 
lens at the output end of the Kr cell, which was positioned to create a collimated VUV beam 
of 4 mm diameter, and the 
$\omega_1$ and $\omega_2$ laser beams came to a focus outside the apparatus. 
The $\omega_1$ laser beam was fixed at 212.55 nm to coincide with the two-photon resonance transition $4p^55p[1/2]_0 \leftarrow 
4p^6$ of Kr \cite{Marangos:JOSAB7:1254}, and the $\omega_2$ laser was tuned in the range of 570-579 nm 
to 
cover 
the wavelength range of the VUV for state-selective excitation of the O($^3P_{j=0,1,2}$) 
photofragments to the \tripS state. The O($^3P_{j=0,1,2}$) products were probed using a two-color 
resonance-enhanced multiphoton ionization (1+1' VUV + UV REMPI) scheme in which the same VUV laser 
beam was used for the excitation step and a $\ordsim 289$~nm laser was used to subsequently ionize 
the atoms. This 1' ionization beam was produced by frequency-doubling a fraction of the $\omega_2$ 
laser beam 
and was counter-propagating to the VUV beam. The size and flux of this $2\omega_2$ beam was adjusted 
for maximum signal
using a 20 cm focal length lens positioned 25-30 cm from the crossing point. The $\omega_2$ laser 
wavelength was scanned by $\pm 4$~cm$^{-1}$ over the nascent O-atom Doppler profile in order to 
uniformly detect all O$^+$ fragments with different velocities. A wavemeter was used to calibrate all 
reported wavelengths.

All lasers were linearly polarized and arranged such that the polarization of the VUV was parallel to 
the plane of the MCP detector, which is the horizontal (H) plane in our setup. For the 1 + 1' REMPI 
detection scheme, the ionizing 
laser polarization was set at 54.7$^\circ$ (magic angle), M, relative to the exciting laser polarization in 
order 
to negate any polarization sensitivity of the ionization step when measuring the 
angular momentum alignment of the nascent O($^3P_{j=0,1,2}$)  products. This set of polarizations is 
labeled from here on as HM. We also compared ionization at the 
magic angle with the ionizing laser polarization set parallel and perpendicular (V) to the detector plane, thus HH 
and HV polarizations, respectively for our apparatus. The detector homogeneity was normalized using 
the geometry VM.

\section{Results}
\begin{figure}
	\centering
	\includegraphics[width=0.45\columnwidth]{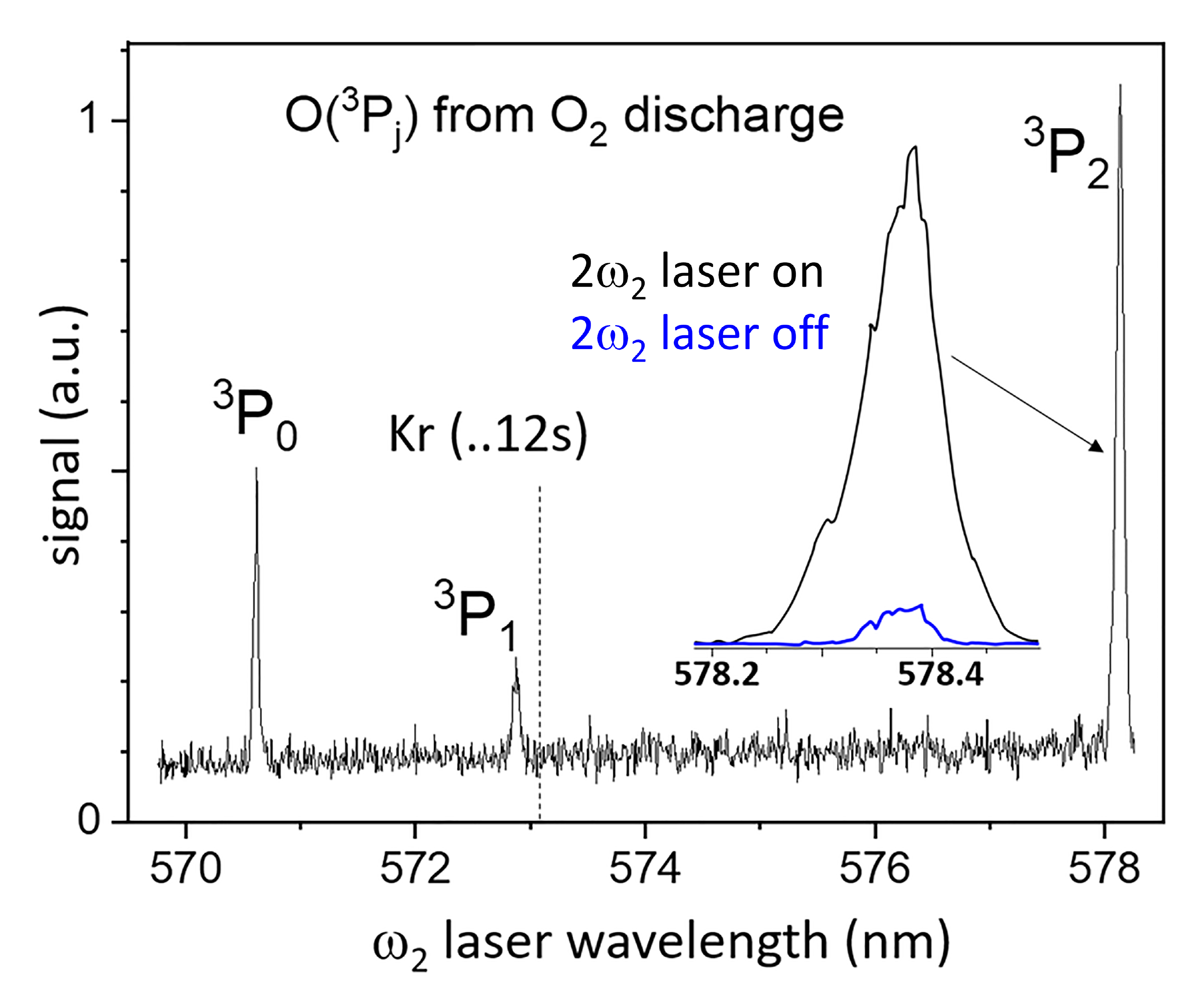}%
	\caption{O$^+$ total ion yield spectrum on scanning the $\omega_2$ laser wavelength to generate 
		tunable VUV in the vicinity of the \tripS resonance of O($^3P_j$) atoms.}
	\label{fig:spectrum}
\end{figure}

A weak resonant O$^+$ signal was obtained when intersecting the O($^3P$) discharge beam \cite{Farooq:PCCP16:3305} with radiation 
from the Kr cell. Under these conditions the ionization step of the 1 + 1' REMPI scheme is provided by 212 
nm radiation, in accord with the results of Ref.~\cite{Lambert:JCP121:10437}. A significant increase 
in O$^+$ signal -- up to 34 times at the highest laser fluence –- was obtained on addition of the 
$2\omega_2$ laser beam. \autoref{fig:spectrum} 
shows the resulting O$^+$ signal on scanning the wavelength of the $\omega_2$ laser from $569-579$~nm. Three 
peaks are observed at the expected positions for resonant excitation of the O($^3P_{j=0,1,2}$) states. The VUV wavelengths 
and energies correspond to O($^3P_2$): 130.214 nm, 9.521 eV, O($^3P_1$): 130.483 nm, 9.502 eV, and 
O($^3P_0$): 130.600 nm, 9.493 eV, in accord with the literature values~\cite{Kramida:NIST}. Off-resonance, a background O$^+$ 
signal is observed due mainly to the 212 nm radiation, as seen by pumping the Kr out of the cell and thereby removing the VUV 
radiation.

The relative intensities of the O($^3P_j$) peaks in \autoref{fig:spectrum} are quite different from
those measured for a similar discharge beam \cite{Farooq:PCCP16:3305} with 2 + 1 REMPI at $\ordsim 226$~nm, or for a microwave 
discharge beam with two-photon resonant degenerate four-wave mixing spectroscopy \cite{Konz:OC134:75}, which show a 
regular decrease in signal on going from $j = 2$ to $j = 1$ to $j = 0$.   
The 1 + 1' REMPI spectrum in \autoref{fig:spectrum}, however, is quite irregular, with a particularly weak $^3P_1$ peak, and  a 
$^3P_2$ peak that is stronger than $^3P_0$. VUV generation in this 130 nm region is affected by 
accidental resonances in Kr at 
2$\omega_1$ + $\omega_2$, see \autoref[b]{fig:schemes}, which increase XUV 
frequency sum generation at the cost of the desired difference frequency VUV. The MgF$_2$ lens at the exit of the Kr cell blocks 
any XUV radiation from the detection chamber. In particular, the allowed $4s^2 4p^5 (^2P^0_{3/2})12s \leftarrow 5p[1/2, 0]$
transition at $\omega_2 = 573.03$~nm appears to most affect the sensitivity for O($^3P_1$). The scan shown in 
\autoref{fig:spectrum} was taken at a Kr pressure of 20 mbar. VUV output at the O($^3P_1$) wavelength can be increased by 
lowering the Kr pressure, but this leads to an overall decrease in signal across the full wavelength range of 
\autoref{fig:spectrum}. Similar effects most likely explain the $^3P_2/^3P_0$ intensity ratio. 

\subsection{Photodissociation of O$_2$ at 130 nm}
The improved 1 + 1' REMPI method was applied to VUV photodissociation of O$_2$ at the three VUV 
wavelengths nearby 130 nm used for resonant detection of O($^3P_j$) fragments. While dissociation limit (DL) 3 
(\autoref[c]{fig:schemes}) is accessible below 133 nm, dissociation is found to take place only to 
limits DL 1 and 2 \cite{Lambert:JCP121:10437}, creating faster and slower O($^3P$) atoms, respectively. Relative 
branching ratios for DL 2/DL 1, product channel angular distributions, and information on $m$ state 
polarization of the products are obtained. The results will be discussed in the following.

\begin{figure}
	\centering
	\includegraphics[width=0.6\columnwidth]{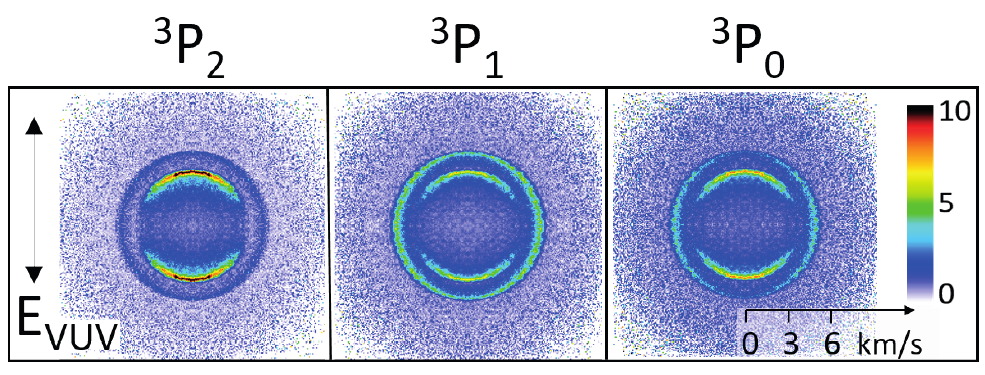}%
	\caption{Velocity mapped O$^+$ images (HM geometry), after symmetrization and correction for the detector inhomogeneity, 
	from O$_2$ photodissociation with VUV radiation tuned to 1 + 1' REMPI of the O($^3P_j$) resonances shown in 
	\autoref{fig:spectrum}. The $\bf{E}$ field direction of the linearly polarized VUV radiation is shown on the left and the 
	color bar on the right codes the signal intensity.}
	\label{fig:images}
\end{figure}

O$^+$ images obtained by crossing an O$_2$ molecular beam with VUV radiation tuned to 
O($^3P_{j=2,1,0}$) resonances are shown in \autoref{fig:images}. The VUV causes both O$_2$ 
photodissociation and O($^3P_j$) excitation as the first step in the 1 + 1' process, where 1' is 
driven by the counter-propagating $2\omega_2$ laser beam. The $\bf{E}$ field direction of 
the linearly polarized VUV ($\bf{E}_{\text{VUV}}$) is indicated in the figure and the polarization direction of the $2\omega_2$ 
ionization laser is set at  \degree{52.4} (magic angle) relative to $\bf{E}_{\text{VUV}}$. 
Two recoil rings are apparent in each image, corresponding to the fast (DL 1) and slow (DL 2) 
dissociation channels. Qualitatively, the fast ring peaks in the horizontal direction, \ie, perpendicular to 
$\bf{E}_{\text{VUV}}$, which indicates a direct (axial) dissociation via a perpendicular ($^3\Sigma^-_g \rightarrow \ ^3\Pi_u$) 
transition in O$_2$. The slower ring peaks along the vertical direction, \ie, along $\bf{E}_{\text{VUV}}$, indicating a 
parallel ($^3\Sigma^-_g \rightarrow \ ^3\Sigma^-_u$) transition.

\begin{figure}
	\centering
	\includegraphics[width=0.45\columnwidth]{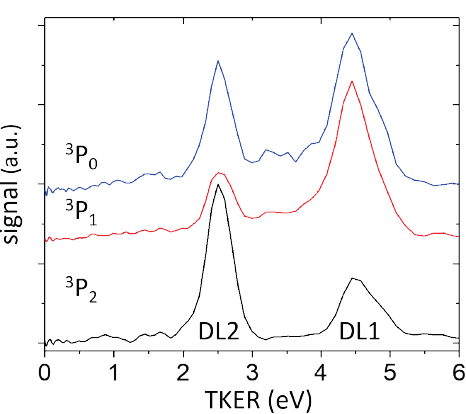}%
	\caption{Total kinetic energy release (TKER) distributions from the images shown in \autoref{fig:images}, 
		after inversion using FINA. 
		The $^3P_0$ and $^3P_1$ curves are given a vertical offset for clarity.}
	\label{fig:TKER}
\end{figure}

A broad background, which is still partially present when the VUV wavelength is tuned outside the Doppler profile of the 
O($^3P_j$) resonances, underlies the recoil rings in each image. It was not possible to 
reliably remove this background directly by any experimental `on-off' subtraction method.   
Furthermore, by observing the regions in the image where the recoil signal is weak or not present, it 
was also apparent that the background varies slightly with the polarization direction of the intense 
ionization laser. For this reason, images taken with the ionization laser polarization set at the 
magic angle were fully analyzed after subtracting a constant value at each pixel, typically $< 5\%$ of 
the maximum intensity. The subtracted value was chosen 
so that no negative intensity appeared across the background-corrected or inverted image. The FINA inversion program 
\cite{Thompson:JCP147:013913} was used to extract the DL 1 and DL 2 angular distributions and the relative branching ratios from 
the 
kinetic energy distributions shown in \autoref{fig:TKER}, despite the possible effect of O-atom 
alignment on the cylindrical symmetry of the images. Integration of each peak in the TKER curves 
yielded the following 
relative branching ratios $\displaystyle \Phi = \frac{\text{DL 2}}{\text{total}}$ at each dissociation wavelength: $\Phi = 0.45 
\pm 
0.05$ 
for O($^3P_0$) at 130.214 nm, $\Phi = 0.28 \pm 0.05$ for O($^3P_1$) at 130.483 nm, and $\Phi = 0.69 
\pm 0.05$ for O($^3P_2$) at 130.600 nm. 

\begin{figure}
	\centering
	\includegraphics[width=0.6\columnwidth]{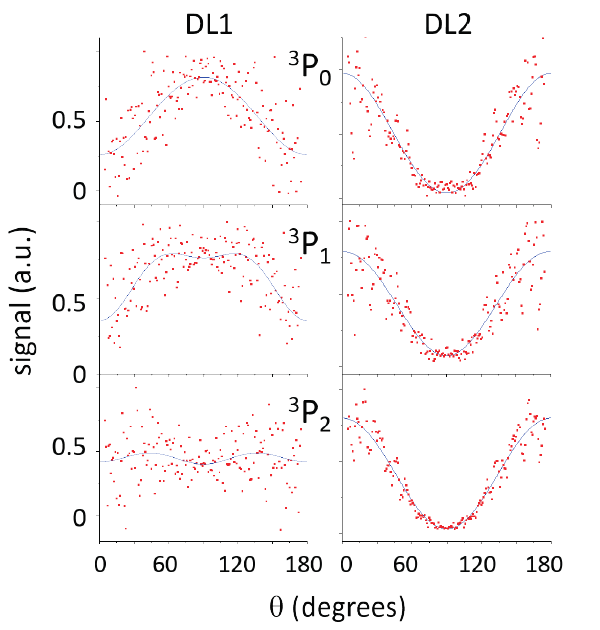}%
	\caption{Angular data (scattered points) and fits (solid lines) for the DL 1 and DL 2 recoil rings obtained from the images 
	with HM polarization shown in \autoref{fig:images}, after inversion using FINA. Values for $\beta^{\mathrm{obs}}$ and 
	$\gamma^{\mathrm{obs}}$ 
	recovered by the fitting 
	routine are listed in \autoref{tab:values}.}
	\label{fig:angdist}
\end{figure}

\begin{figure}
	\centering
	\includegraphics[width=0.45\columnwidth]{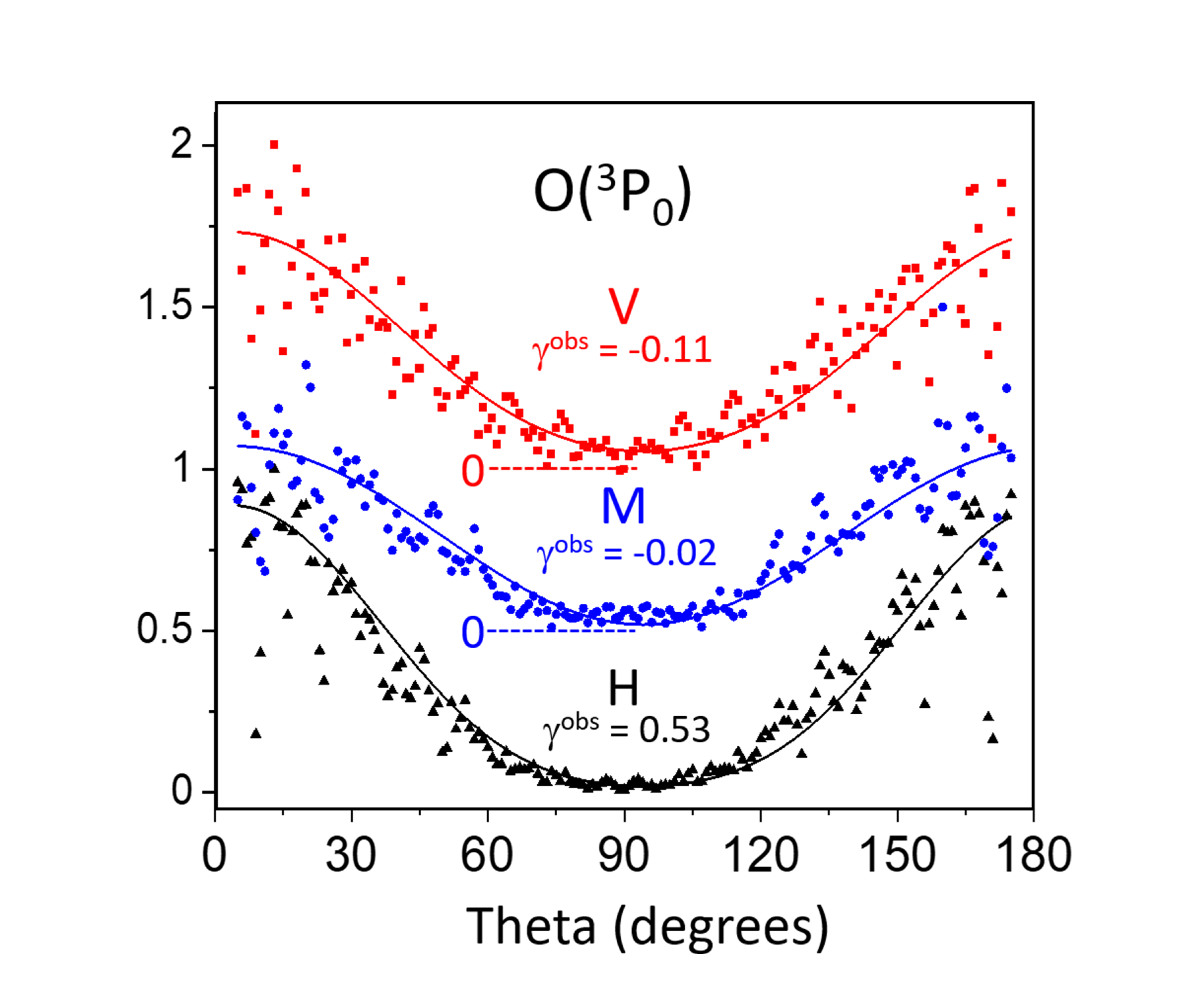}%
	\caption{Angular data (scattered points) and fits (solid lines) for the DL 2 recoil ring obtained 
		from the O($^3P_0$) images with HV (red), HM (blue), and HH (black) polarization. Values for $\gamma^{\mathrm{obs}}$ 
		from the 
		fitting routine are indicated for each curve. The signals for HV and HM polarization are given a vertical offset for 
		clarity.}
	\label{fig:angdist2}
\end{figure}

Angular distributions for the recoil rings in the inverted images are plotted in 
\autoref{fig:angdist}. Angular distributions from O($^3P_0$) DL 2 using the same background subtraction correction and HV, HM, 
and HH polarizations are shown in \autoref{fig:angdist2}.
The observed angular distributions are described by the equation \cite{Lambert:JCP121:10437}
\begin{equation}
	I(\theta) = C_0 (1 + \beta^{\mathrm{obs}} P_2 (\cos\theta) + \gamma^{\mathrm{obs}} P_4(\cos\theta)),
\end{equation}
where $P_n (\cos\theta)$ are Legendre polynomials. This equation was used for fitting the image data 
(scattered points) in \autoref{fig:angdist} and \autoref{fig:angdist2}. 
Best fitting coefficients for $\beta^{\mathrm{obs}}$ and $\gamma^{\mathrm{obs}}$ are listed in \autoref{tab:values}. 
As discussed in more detail later, a non-zero value for $\gamma^{\mathrm{obs}}$ is an indication of alignment effects arising 
from the 
REMPI process for O($^3P_j$) detection. 

\begin{table*}[t]
	\centering
	\caption{Anisotropy parameters $\beta$ and $\gamma$, and fractional populations $A_\lambda$ of the 
		different $\lambda$ levels, where $A_{-\lambda} = A_{\lambda}$, for the different O($^3P_j$) states for the two 
		different dissociation limits (DL).}
	\label{tab:values}
	\begin{tabular}{c c | D{.}{.}{1} D{.}{.}{2} D{.}{.}{2} c c c}
		State & DL & \multicolumn{1}{c}{$\beta^{\mathrm{obs}}$} & \multicolumn{1}{c}{$\gamma^{\mathrm{obs}}$} & 
		\multicolumn{1}{c}{$\beta$} & \multicolumn{1}{c}{$A_0$} & \multicolumn{1}{c}{$A_1$} & \multicolumn{1}{c}{$A_2$} \\  
		\hline
		$^3P_0$ & 2 & 1.8 & 0.04 & 1.8 & 1 & - & - \\
		$^3P_0$ & 1 & -0.6 & -0.02 & -0.6 & 1 & - & - \\
		$^3P_1$ & 2 & 1.5 & -0.06 & 1.6 & 0.4 & 0.3 & -  \\
		$^3P_1$ & 1 & -0.3 & -0.2 & 0.5 & 0.8 & 0.1 & - \\
		$^3P_2$ & 2 & 1.8 & -0.02 & 1.8 & 0.0 - 0.5 & 0.3 - 0.0 & 0.2 - 0.3 \\
		$^3P_2$ & 1 & 0.1 & -0.2 & 0.7 & 0.0 - 0.2 & 0.2 - 0.0 & 0.3 - 0.4 \\
		&   &     &      & -0.5 & 0.4 - 0.8 & 0.3 - 0.0 & 0.0 - 0.1
	\end{tabular}
\end{table*}

\subsection{Analysis}
Molecular dissociation by polarized radiation can produce polarized atomic fragments, and analysis of 
this nascent atom polarization affords unique insights into the dissociation process \cite{Brunt:JCP48:4304, Clark:PCCP48:5591}. 
In the case of molecular oxygen, for example, photoexcitation from the upper $^3\Sigma^-_u$ 
state of the Schumann-Runge continuum leads to 89\% of the produced O($^1D$) atoms in the $m=0$ state 
\cite{Wu:MP108:1145, Lambert:JCP121:10437, Eppink:JCP108:1305, Lewis:JCP110:11129}, which can be understood based on the 
long-range quadrupole-quadrupole interaction \cite{Wu:MP108:1145}. Assuming that the electron spin is not polarized, the 
$|m|$-populations are the same for both atoms in a $\Sigma$ state. The co-partner O($^3P$) atoms are therefore expected 
to show similar $m$-state polarization.  
The strong alignment of nascent O($^1D$) atom products results in their ionization probability 
depending significantly on the recoil direction in the laboratory (laser polarization) frame, i.e., 
the expected angular distribution is perturbed by the atomic alignment.  Unraveling nascent atom 
polarization information from photofragment imaging data is a non-trivial task which is dependent on 
the details of the REMPI process used \cite{Suits:CR108:3706}. A full polarization study requires 
using linearly and circularly polarized laser beams in several pump-probe geometries.  We use here only 
linear laser polarization and a semi-classical analysis where coherence effects, which can be 
substantial for O($^3P$) atoms \cite{Brouard:JPCA108:7965, Gilchrist:JCP138:214307}, are not probed.  
Our method provides two-dimensional angle-speed distributions of O$^+$ ions, while the strategy is to 
determine alignment information based on the known alignment sensitivity of the O($^3P$)+ $h\nu_{\text{VUV}} 
\rightarrow$ O(\tripS) transition.  

In general, an experimentally observed photofragment angular distribution can be described by the 
following function \cite{Lambert:JCP121:10437}
\begin{equation}
	I(\theta) = C \left[ 1 + \beta P_2 (\cos \theta) \right] \sum_\lambda A_\lambda F_\lambda(\theta).
\end{equation}
Here, $\beta$ is the spatial anisotropy parameter describing the angular distribution of the 
photodissociation fragments, $\theta$ is the angle between the dissociating laser polarization and the recoil velocity, $P_2$ is 
a Legendre polynomial of order 2, $\lambda$ is the projection of the angular 
momentum $j$ on the recoil direction, $A_\lambda$ is the fractional population of each $\lambda$ level, and $F_\lambda 
(\theta)$ is the probe frame angular detectivity function, which depends on the transition probed.
When $F_\lambda(\theta)$ is known, the alignment-free $\beta$ parameter and the $\lambda$ populations $A_\lambda$ can, in 
principle, be 
determined from $\beta^{\mathrm{obs}}$ and $\gamma^{\mathrm{obs}}$ for each $j$-state.  In 
\cite{Vroonhoven:JCP116:1965}, $F_\lambda(\theta)$ equations were derived for two-photon excitation of 
O($^3P$) to the $3p \ ^3P$ state where the intermediate state was assumed to be only the \tripS 
state.  The same derivation becomes rigorous with our 1 + 1' REMPI scheme, where the upper state 
becomes the ion continuum with $j = 0, 1, 2$ allowed to be populated after ionization. In 
Ref. \cite{Vroonhoven:JCP116:1965}, the 
ionization step was assumed to be saturated, \ie, only the resonant transition is polarization 
sensitive. We can vary the polarization direction of the $2\omega_2$ ionization laser beam as a check 
of the ionization step 
saturation. The H polarized VUV laser beam drives both photodissociation and the resonant excitation step of the O($^3P_j$) 
ionization scheme. In the O$_2$ 
photoexcitation step, a subset of the randomly oriented O$_2$ molecules that are 
	pointing primarily along (for a parallel $\Sigma \rightarrow \Sigma$ transition) or perpendicular (for a $\Sigma 
	\rightarrow \Pi$ transition)
	to the $\bf{E}$ field direction of the linearly polarized VUV beam 
	is photo-selected, and their subsequent direct (axial) dissociation creates a product angular 
	distribution described by $\beta$. Photodissociation also creates O-atom fragments that are aligned in the molecular frame 
	with 
respect to the interatomic bond axis.  Only those molecules pointing directly 
along $\bf{E}$ create fragment atoms traveling in the $\bf{E}$ direction, and these atoms are excited 
to the \tripS state following the selection rule $\Delta m=0$, with $m$ the projection of $j$ on the laser polarization 
direction. Along this direction, the $m=2$ state of $^3P_2$ cannot be 
excited, and when starting from the $^3P_0$ state, the intermediate state becomes aligned since only the $m=0$ 
level of the 
\tripS upper state is populated. 
When $\beta \sim 2$ as for the DL 2 channel, 2 + 1 REMPI at 226 nm and 1 + 1' REMPI via the \tripS state 
are less sensitive for $m=2$ detection, as pointed out by Lambert \etal \cite{Lambert:JCP121:10437}. 
Most excited molecules are not pointing directly along $\bf{E}$, and the recoil frame is thus rotated from 
the molecular frame. The projection of $m$ onto the recoil 
frame results in mixing of $m$ 
levels and thus excitation to the $\lambda = \pm 1$ of the \tripS atom.  Similar 
arguments provide an excitation pathway for the $\lambda=2$ state of O($^3P_2$) via \tripS to $\lambda=2$ in 
the ion continuum. In the recoil frame, all $\lambda$ states can thus be detected.

Assuming that the nascent oxygen atoms are in a specific $j,\lambda$ fine structure state, the probe frame 
angular 
detectivity functions for a two-photon transition are given by \cite{Vroonhoven:JCP116:1965}
\begin{equation}
	F_\lambda(\theta) = 1 + \alpha_\lambda P_2(\cos \theta),
\end{equation}
with 
\begin{equation}
	\alpha_\lambda = \frac{\rho_0^{(2)}(j,\lambda)}{\rho_0^{(0)}(j,\lambda)}I_2(j).
\end{equation}
Here, $I_2(j)$ is a geometrical factor \cite{Vroonhoven:JCP116:1965}, and $\rho_0^{(k)}(j)$ 
are density matrix moments, which are given by
\begin{equation}
	\rho_0^{(k)}(j) = \sum_{\lambda=-j}^{j} (-1)^{j-\lambda} \braket{j, \lambda, j, -\lambda | k, 0} p_{j,\lambda},
\end{equation}
with $p_{j,\lambda}$ the population of the $j,\lambda$ state. For a pure $j,\lambda$ state, this results in 
\begin{equation}
	\frac{\rho_0^{(2)}(j,\lambda)}{\rho_0^{(0)}(j,\lambda)} = \frac{\braket{j, \lambda, j, -\lambda | 2, 0}}{\braket{j, \lambda, 
	j, 
	-\lambda | 0, 0}}.
\end{equation}
For the O($^3P_j$) system we find that $\alpha_0 = 1$, $\alpha_1 = 1/2$ and $\alpha_2 = -1$ for $j = 
2$, $\alpha_0 = -1$ and $\alpha_1 = 1/2$ for $j = 1$, and $\alpha_0 = 0$ for $j = 0$ for the \tripS 
intermediate state~\cite{Vroonhoven:JCP116:1965}. Altogether, this results in analytical expressions for $\beta^{\mathrm{obs}}$ 
and $\gamma^{\mathrm{obs}}$ in terms of $\beta$, $\alpha_\lambda$, and $A_\lambda$:
\begin{equation}
	\begin{aligned}
		\beta^{\mathrm{obs}} &= \frac{\sum_{\lambda=-j}^j A_\lambda \left[ \beta + \alpha_\lambda + \frac{2}{7}\beta 
			\alpha_\lambda \right]}{\sum_{\lambda=-j}^j A_\lambda (1 + \frac{1}{5}\beta \alpha_\lambda)}\\
		\gamma^{\mathrm{obs}} &= \frac{\sum_{\lambda=-j}^j \frac{18}{35} A_\lambda \beta \alpha_\lambda}{\sum_{\lambda=-j}^j 
		A_\lambda (1 + \frac{1}{5}\beta \alpha_\lambda)}.
	\end{aligned}
\end{equation}
From these expressions, in combination with 
\begin{equation}
	\sum_{\lambda=-j}^{j} A_\lambda = 1,
\end{equation}
we can analytically determine $\beta$ and the $A_\lambda$ fractional populations from $\beta^{\mathrm{obs}}$ 
and $\gamma^{\mathrm{obs}}$. 

\autoref{tab:values} contains the resulting $\beta$ and $A_\lambda$ values we 
found for all images. In the case of O($^3P_2$) there are too many parameters to obtain one unique solution. We 
therefore determined the ranges of possible fractional populations. We obtain one solution with a positive $\beta$, and one with 
a negative $\beta$ value. For DL 2, we only listed the solution for the positive $\beta$ value, since $\beta$ is also positive 
for the other two $j$ states. For O($^3P_2$) DL 1, for which the signal-to-noise ratio in the experimental data is low 
(\autoref{fig:angdist}), it is unclear which solution is more probable and we therefore listed both solutions.

General trends are that for DL 2, $\beta$ is 
close to 2, indicating that the 
transition is close to parallel. The observed anisotropy parameter for DL 1 is smaller than 1 and even negative for O($^3P_{j 
= 0}$), which 
indicates that this transition is more perpendicular. The most significant alignment effects, based on $\gamma^{\mathrm{obs}}$, 
appear in 
the DL 1 channel where for O($^3P_1$) DL 1 the most population is found in the $|\lambda| = 0$ state. 

\section{Discussion}
1 + 1' REMPI detection of O($^3P$) atoms using resonant VUV radiation around 130 nm has been 
described here as an alternative to standard 2 + 1 REMPI at 226 nm.  By applying intense $2\omega_2$ 
radiation for the ionizing step, the sensitivity of 1 + 1' detection is greatly increased (see the 
inset of 
\autoref{fig:spectrum}), 
and this approach provides much lower ion recoil, \ie, 9 m/s instead of 34 m/s for 2 + 1 REMPI. While recoil is not 
an issue in this or most other 
photodissociation experiments, the separation of the resonant excitation and the ionization step does 
allow probing the polarization sensitivity of the method. In terms of relative fine structure yields, 
the interfering resonances for VUV generation in the 130 nm region made a measurement of the total 
$j$-dependent O($^3P_j$) relative yields unreliable in this study.  This problem could be addressed by 
using a secondary calibration method such as 2 + 1 REMPI under the same conditions. 

1 + 1' REMPI should be competitive with 2 + 1 REMPI for the following reasons.  The resonant \tripS 
state of O($^3P$) is the lowest optically allowed excited state, with a lifetime of $\ordsim 1.8$~ns \cite{Tayal:PS79:015303}
and, for excitation from the ground state, a maximum absorption cross section $\sigma = 7.3 \times 10^{-14} \ \text{cm}^2$, 
assuming the lineshape of a Doppler profile at 300 K \cite{Parkes:AJ149:217}. We estimate our VUV production at roughly $N_{VUV} 
= 10^{10}$ photons/pulse \cite{Hanna:IJMS279:134} focused to $A \ordsim 0.1$~cm$^2$, which for a pulse length $\tau = 4$~ns 
yields a rather large excitation probability of $P = \sigma I \tau \ \ordsim \ 0.07$, with $I = N_{VUV}/(A\tau)$.
For both 1 + 1' and 2 + 1 REMPI, the ionization rate is usually smaller than the resonant excitation rate.  
However, when using strongly focused 226 nm radiation, the ionization step is expected to 
`saturate', and thus not affect the polarization sensitivity. In this study with 1 + 1' REMPI, the 
intense (2.6 mJ/pulse) $2\omega_2$ pulse is also well-focused in order to drive ionization during the 
4 ns 
pulse length, but the O($^3P_0$) polarization dependences (\autoref{fig:angdist2}) show that 
saturation is clearly not reached. 
While the VUV flux is relatively low, the high-energy photons create 
background at the mass/charge ratio of O$^+$, as observed in \autoref{fig:spectrum}.  
The intense UV radiation used for O($^3P$) 2 + 1 REMPI at 226 nm, however, can cause similar effects.  
For example, more O($^3P$) signal usually arises from two-photon (VUV equivalent) dissociation of 
O$_2$ than from one-photon UV dissociation~\cite{Buijsse:JCP108:1998}.   

In future studies, 
the use of a third dye laser for the 1' ionization step could have several advantages. First of all, 
the ion recoil could be reduced even further by tuning the wavelength of this laser to ionize just 
above the threshold. Second, a common state for exciting and the same wavelength for ionizing the 
different fine structure states could be used, making the REMPI spectra easier to interpret. 
Additionally, the 1' ionization wavelength could then be tuned to reach an autoionizing Rydberg state 
to possibly obtain a higher ionization efficiency while maintaining the low ion recoil.

\subsection{Photodissociation of O$_2$ in the 130 nm region}
\begin{figure}
	\centering
	\includegraphics[width=0.45\columnwidth]{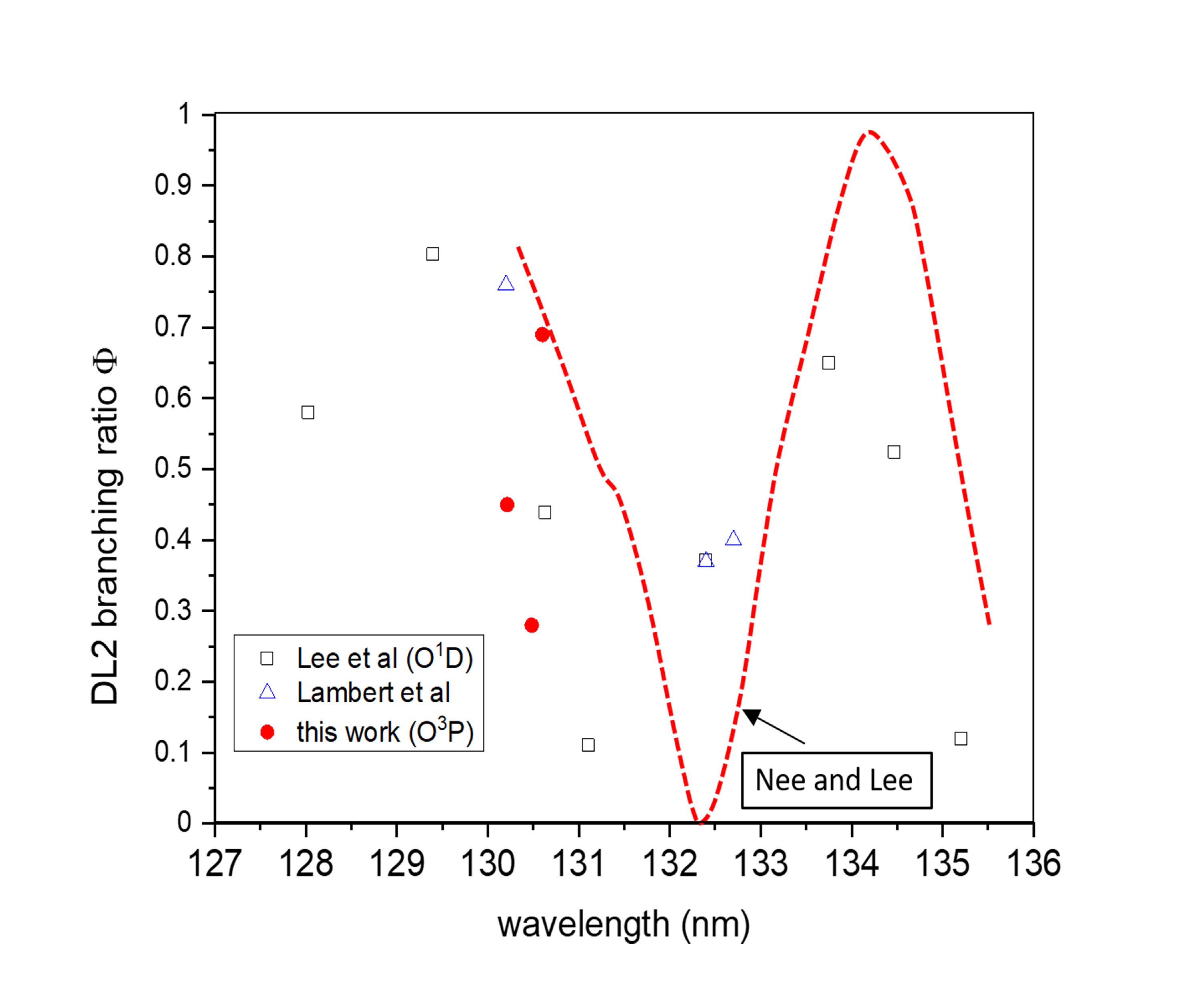}%
	\caption{DL 2/total branching ratios in the 127 - 136 nm region. Closed circles are from the 
		present work with O($^3P_j$) detection (assuming a statistical $j$-state distribution), open 
		triangles from \cite{Lambert:JCP121:10437}, open squares from Ref.~\cite{Lee:JCP67:5602}, and the continuous 
		dashed-line is from Ref.~\cite{Nee:JPCA101:6653}.} 
	\label{fig:branching}
\end{figure}

O$_2$ absorption in the range from 130-137 nm is most interesting as the only spectral region where 
two allowed transitions, \ie, the high energy tail of the $X \rightarrow B$ Schumann-Runge continuum (SRC), 
and $X \rightarrow ^3\Pi_u$, are known to overlap with similar intensity \cite{Balakrishnan:JCP112:1255, 
	Lewis:PCE26:519}.  Both upper states 
in these transitions undergo avoided crossings with Rydberg states of the 
same symmetry, causing shoulders in the repulsive walls of each potential energy curve, which are 
challenging to quantify accurately.   Adiabatic correlation diagrams connect the $B$ state to 
(O($^3P_2$) + O($^1D$)) of DL 2, and the $^3\Pi_u$ state to (O($^3P_2$) + O($^3P_1$)) of DL 1.  Advanced 
theory \cite{Allison:JGR87:923, Gibson:JESRP80:9} predicts interference effects resulting from overlap, where the 
maximal 
contributions by the weaker $^3\Pi_u$ state peak around 135.1 and 131.5 nm.  Qualitatively, it can thus be 
expected that the DL 2 branching ratios will show a strong dip at these wavelengths.  Our branching 
ratio data is quite limited in that single O($^3P_j$) fine structure states are detected, each at a 
different dissociation wavelength near 130 nm.  Furthermore, due to the varying strength of the 
generated VUV, it was not possible to directly calibrate the $j$-state yields.  Measurements by 
Lambert \etal \cite{Lambert:JCP121:10437} in this spectral region suggest that the relative O($^3P_j$) fine 
structure yields are statistical, \ie, the population ratios for $j = 2:1:0$ are equal to $5:3:1$, respectively, for the
O($^3P_j$) products from both DL 1 and DL 2. This can be expected given the large kinetic energy releases for both 
channels. Assuming this to be the case, the relative DL 2/total branching ratios we measure can be 
compared with the O($^1D$) branching ratios measured previously, which are shown in 
\autoref{fig:branching}. 

Our data and that of Ref.~\cite{Lambert:JCP121:10437} are from a cold molecular beam of O$_2$ and narrow 
band lasers, while Refs.~\cite{Lee:JCP67:5602, Nee:JPCA101:6653} employed samples of O$_2$ in a gas cell.  A lamp with 
monochromator light source was used in \cite{Lee:JCP67:5602}, while data from Nee and Lee were 
obtained using a continuously tunable VUV synchrotron source \cite{Nee:JPCA101:6653}. While the scatter in experimental data is 
significant, due to the 
different light source bandwidths and/or discrete points reported, the 
	overall agreement is reasonable. It is apparent especially from our own data that the branching ratios oscillate rapidly 
in this wavelength region, on a very fine wavelength scale.   

In general, the alignment-free angular anisotropy parameters $\beta$ extracted from our data are also consistent 
with the states involved in the adiabatic correlation diagram, where excitation to the $B$-state leads to $\beta \ordsim 1.8$ 
for DL 2 and excitation to the $^3\Pi_u$ state leads to $\beta \ordsim -0.6$ for O($^3P_0$) DL 1, see 
\autoref{tab:values}. These measured values suggest that curve crossing after optical excitation (which would lead to less 
extreme values of $\beta$ ($-1< \beta < 2$)), is present but minor. The anisotropy parameter for O($^3P_1$) DL 1 suggests that 
curve crossing here has a larger effect. While Lambert \etal did not report $\beta$ values at 130.2 
nm or 
132.7 nm, their O($^3P_2$) image at 132.7 nm shows quite similar features as our images shown in 
\autoref{fig:images}.  

Our $A_\lambda$ population values presented in \autoref{tab:values} are less certain than desired and 
will also be affected by the curve crossing mentioned above. Lambert \etal \cite{Lambert:JCP121:10437} did 
not report alignment information at 130.2 nm or 132.7 nm, and their results for the strong absorption 
peaks around 124 and 120 nm could be affected by a polarization dependence of the ionization step 
(their laser polarization corresponds to our HH in \autoref{fig:angdist2}).  In general, their 
analysis indicated a dominant $A_{\lambda=0}$ component for DL 2, which is not so clear from our data. A dominant 
$A_{\lambda=0}$ component was found for nascent O($^1D$) fragments from O$_2$ photodissociation to DL 2 in the 120-125 nm 
region, as well as across the $X \rightarrow B$ Schumann-Runge continuum\cite{Lambert:JCP121:10437, Eppink:JCP108:1305, 
Wu:MP108:1145, Lewis:JCP110:11129}. Previous measurements in the 140-165 nm region did not detect any significant alignment of 
the O($^3P_{1,2}$) products for DL 2~\cite{Wu:MP108:1145}. 
At present there are no simple predictions for alignment of DL 1 fragment atoms produced via $^3\Pi_u$
excitation. Our data for O($^3P_1$) DL 1 suggests a dominant $A_{\lambda=0}$ component.    

Our results can moreover be compared to predictions by adiabatic correlation diagrams. For DL 1, the $^3\Pi_u$ states 
correlates with the production of O($^3P_2$) + O($^3P_1$), whereas for DL 2, the $B$-state is correlated with O($^3P_2$) + 
O($^1D_2$) \cite{Huang:JCP94:2640}. The adiabatic correlation diagrams therefore predict that no O($^3P_0$) will be formed for 
both dissociation limits, 
and that no O($^3P_1$) will be formed for DL 1. Since our results are not in agreement with these simple predictions, this is 
another indication that curve crossing after optical excitation and contributions from Rydberg states play a role at the high 
photon energies used in this study.

The combination of our 1 + 1' REMPI method with a second VUV source tuned 
in small steps across the 130-137 nm region should lead to a more quantitative understanding of the 
allowed VUV transitions of O$_2$. 

\section{Conclusions \& outlook}
We have presented a novel 1 + 1' near-threshold REMPI scheme for the detection of O($^3P_j$) atoms. This scheme 
induces a recoil of only 9 m/s to the O$^+$ ion, instead of 34 m/s for the generally used 2 + 1 REMPI scheme. By using 302 nm 
radiation from a third dye laser for the 1' ionization step, it should even be possible to approach zero recoil. 
Our scheme is more sensitive than the 1 + 1' scheme from Lambert \etal \cite{Lambert:JCP121:10437}, which uses 212 nm radiation 
for the 1' step, and therefore results in higher signal levels. It can moreover be used to extract polarization information for 
the O($^3P$) atoms, as demonstrated by our investigation of the photodissociation of O$_2$ around $\ordsim 130$~nm.

The low ion recoil in combination with the good sensitivity of our scheme provides good prospects for applying this REMPI scheme 
to molecular scattering studies in crossed beam experiments combining controlled molecular beams and VMI. In these studies, the 
ion recoil is one of the limiting factors for the resolution that can be obtained. Moreover, signal levels are generally low, 
such that a high sensitivity is required. Scattering processes of interest include inelastic O($^3P$) + He / H$_2$ 
and reactive C($^3P$) + NO($X^2\Pi$) $\rightarrow$ CN($X^2\Sigma^+$) + O($^3P$) collisions, which are both considered as 
important processes for astrochemistry.

\section*{Acknowledgements}

This work receives funding from the European Research Council (ERC) under the European Union's Horizon 
2020 Research and 
Innovation Program (Grant Agreement No. 817947 FICOMOL) and from the European Union's Horizon 2020 
Research and Innovation Program under the Marie Sklodowska-Curie grant agreement No. 886046 and 
889328. 
The authors thank Niek 
Janssen and Andr\'e van Roij for expert technical support and Xingan Wang, Vikram Plomp, and Zhong-Fa Sun for fruitful 
discussions.

\section*{Disclosure statement}

The authors declare no competing interests.

\end{document}